\documentclass[twoside,english,5p]{elsarticle}
\usepackage[T1]{fontenc}
\usepackage[latin9]{inputenc}
\pdfoutput=1
\usepackage{varioref}
\usepackage{graphicx}
\usepackage{subscript}

\makeatletter

\usepackage{braket}

\makeatother

\usepackage{babel}
\begin{document}

\title{Faster Communication Using Probabilistic Swapped-Bell-States Analysis}

\author{Hauke~Traulsen}

\ead{hauke.traulsen@iis.fraunhofer.de}

\address{Computer Scientist at The Fraunhofer Institute for Integrated Circuits
IIS, Nordostpark 93, 90411 Nuremberg, Germany \\
\medskip{}
This article was written by the author outside of his regular charges.}
\begin{abstract}
In this article, a procedure called Probabilistic Swapped-Bell-States
Analysis (PSBA) is proposed. Using this procedure two communication
partners can transmit (binary-encoded) information over large spatial
distances. This procedure is unusual insofar as no classical communication
channels are used either during or after information encoding. To
make this possible, entanglement swapping is used as a transport channel.
The encoding of bits is realized by the execution (or non-execution,
as the case may be) of entanglement swapping on multiple photon pairs,
while decoding is realized by a statistical detection of swapped entanglements.
If the PSBA procedure sustains itself against a rebuttal by the scientific
community, it would constitute a technical approach for transmitting
information faster than light. The PSBA procedure seems to be in harmony
with the no-communication theorem, since PSBA does not manipulate
(and teleport) specific states from one (entangled) particle to another
in order to communicate, but instead uses statistics on entanglement
swappings for information encoding. 

\medskip{}
\end{abstract}
\maketitle

\section{Introduction}

Entanglement is one of the fundamental concepts of quantum mechanics.
Albert Einstein called this kind of correlation \textquoteleft{}spooky'
\citep{Einstein.1935}), as it can be shown that manipulations on
one constituent of an entangled pair of particles affect the other
constituent very fast: Considering an unknown kind of interaction
between both particles, this transmission would have to be executed
at a speed at least 10,000 times faster than light \citep{Salart.2008}.
There would be unprecedented opportunities if this \textquoteleft{}instantaneous'
influence could be used for sending messages. However, as the no-communication
theorem explains, this kind of influence cannot be used for communication
\citep{Peres.2004}. Fortunately, the no-communication theorem allows
for the detection of whether two photons are entangled or not, and
it does not prohibit the creation and destruction of entanglements
by entanglement swapping. With these thoughts we turn towards the
basic concept of PSBA.

\section{Basic Concept}

With the well-known procedure called \textquotedblleft{}entanglement
swapping\textquotedblright{} it is possible to entangle two photons,
even though they have never interacted with each other \citep{Zukowski.1993,Pan.1998}.
The concept of a special variant of entangle swapping, taken from
\citep{XiaosongMa.2012}, is shown in Figure \vref{fig:concept-of-a-time-delayed}:
A pair of polarization-entangled photons (1\&2) is produced at Alice\textquoteright{}s
location; another pair (3\&4) is produced at Bob\textquoteright{}s
location. Alice sends one photon (2) to Victor. Bob does the same
(photon 3). Alice and Bob measure the polarization of the photons
1 (Alice) and 4 (Bob). Victor decides randomly whether he measures
the polarization of his photons (2\&3) by a separable-state measurement
(SSM) or by a Bell-state measurement (BSM). If he uses the Bell-state
measurement, the entanglements of 1\&2 and 3\&4 will be eliminated,
and new entanglements of the photons (2\&3) and (1\&4) will be created.
As shown in Figure \vref{fig:concept-of-a-time-delayed} and shown
and described in \citep{Ma.2012}, an entanglement swapping can occur
even if the measurements of Alice and Bob occur \emph{before} Victor
makes his decision. But this \textquoteleft{}back-in-time\textquoteright{}
effect is not the main subject of this article. What we need in this
article is the basic mechanism of entanglement swapping, but we will
still discuss this delayed-choice variant of entanglement swapping
later in this article.

\begin{figure}
\includegraphics[width=1\columnwidth]{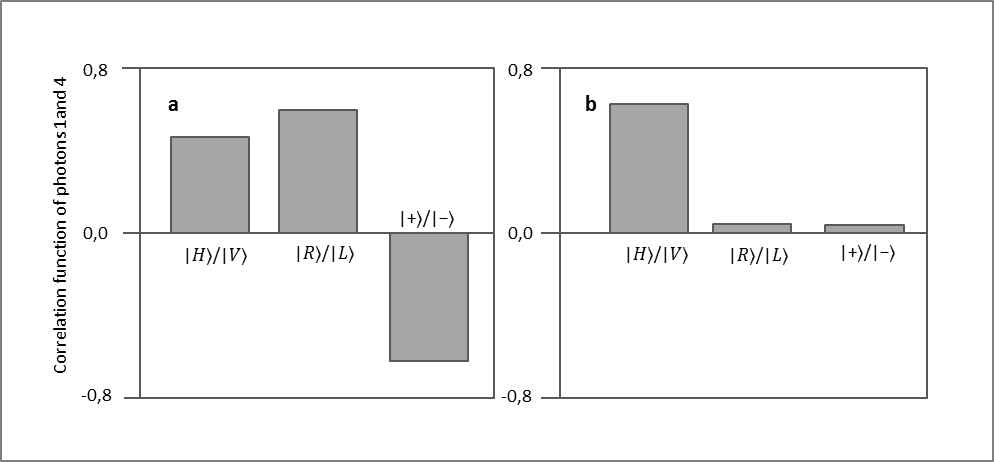}

\caption{Sketch of Figure 3 in \citep{XiaosongMa.2012}: signatures \emph{a}
and \emph{b} for correlations of photons 1 and 4. Situation for\emph{
a}: Victor used Bell-state measurement. Situation for\emph{ b}: Victor
used separable-state measurement. To generate these diagrams measurement
data from Victor, Alice and Bob is needed, but - under certain conditions
(see text) - distinguishable diagrams can be generated without additional
classical communication between Victor and (Alice/Bob).\label{fig:Sketch-of-Figure}}
\end{figure}

\begin{figure}
\includegraphics[width=0.99\columnwidth]{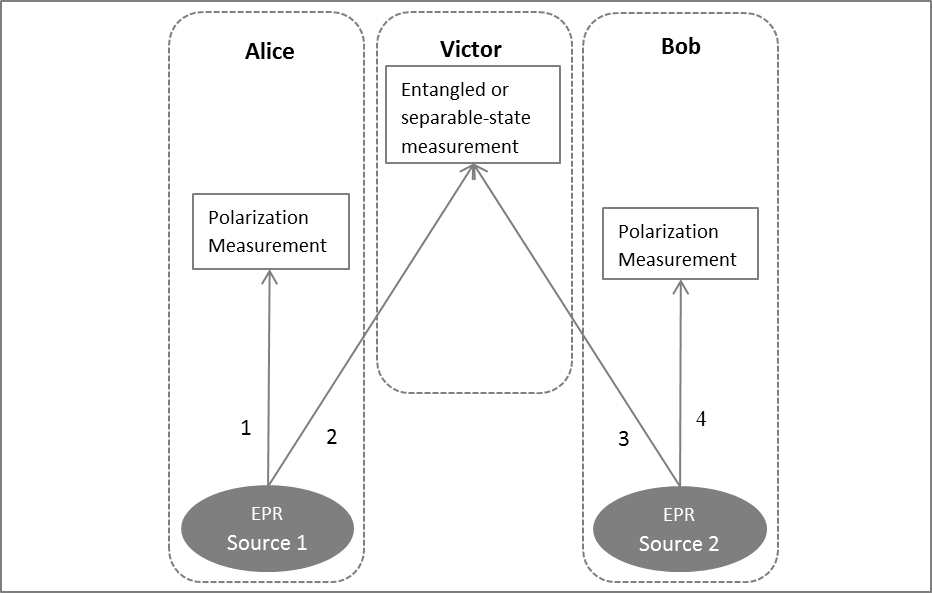}\caption{Concept of a time-delayed entanglement swapping experiment, taken
from\citep{Ma.2012}\label{fig:concept-of-a-time-delayed}}
\end{figure}

This article is inspired by two experimental result diagrams shown
in Figure 3 of \citep{XiaosongMa.2012}. A sketch of these diagrams
is shown in Figure \vref{fig:Sketch-of-Figure}. 
As in the delayed-choice experiment described, Victor\textquoteright{}s
decisions (BSM or SSM) were completely random. This randomness implies
a required sorting (based on Victor's choices) of Alice\textquoteright{}s
and Bob's detected correlations into at least two subsets: correlations
for Victor's BSM (left part of the diagram) and correlations for Victor's
separable-state measurement (right part of the diagram). The author
asked himself (as those two diagrams look quite different): If it's
possible to generate these two diagrams by sorting data produced by
random decision, would it not be possible to generate these diagrams
deliberately? What would happen if Victor decides not randomly, but
deliberately for BSM and produces entanglement swappings (let's say,
for example, 300 times in sequence)? Would we see something like the
left diagram in Figure \vref{fig:Sketch-of-Figure}? And if Victor
decides (again 300 times in sequence) to measure the photons individually
and we count coincidences as described before: Would we see something
like the right diagram? The author assumes that this would be the
case and the rest of this article is based upon this assumption. If
we detect something like \textquoteleft{}the left diagram' (entanglement
swappings), we want to interpret this as a sent binary 1 (1b). If
we detect something like \textquoteleft{}the right diagram' (SSM),
we want to interpret this as a binary 0 (0b). In this article, we
will use an approach for distinguishing a sequence of maximally polarization-entangled
photons and a sequence of unentangled photons without necessarily
distinguishing any of the Bell states unambiguously. It will be shown
that no information from Victor is required to decide probabilistically
whether a sequence of photon pairs (shared by Alice and Bob) is either
maximally entangled or unentangled. 

\begin{figure}
\includegraphics[width=1\columnwidth]{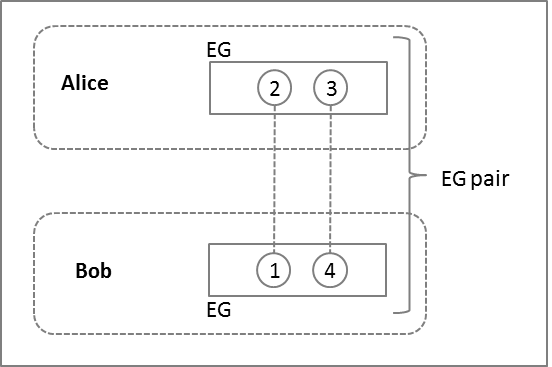}

\caption{Basic concept of entangle groups (EG): A pair of entangled EGs: EG\textsubscript{\emph{A}}
(at Alice) contains photons 2 and 3 which are maximally polarization-entangled
with photons 1 and 4 in EG\textsubscript{\emph{B}} (at Bob). The
entanglement is shown by the dotted lines between the photons.\label{fig:basic-concept-of EMG}}
\end{figure}

But let us go one step back and construct the PSBA concept from the
bottom up. First of all, the suggested PSBA procedure should be interpreted
as a gedankenexperiment due to a lack of technical capabilities to
transport entangled photons over long spatial distances and to store
entangled photons over a longer period of time, whereas proof-of-concept
experiments should be realizable today by using existing technologies.
The latter assumption is reasonable, since the setup of such a proof-of-concept
experiment will have significant similarity to experiments already
realized \citep{Pan.1998,XiaosongMa.2012,Schmid.2009,Weihs.2001}.
We start the setup of this gedankenexperiment by creating two pairs
of maximally polarization-entangled photons (1,2) and (3,4) as also
shown in Figure\vref{fig:concept-of-a-time-delayed}. To generate
these photon pairs, Spontaneous Parametric Down Conversion \citep{Kwiat.1995}
can be used. In contrast to the system architecture assumed for Figure
\vref{fig:concept-of-a-time-delayed} in the PSBA concept,\emph{ both}
photons 1 and 4 are sent to Bob, while Alice gets the photons 2 and
3. We assume Alice and Bob can store their entangled photons until
they are needed for communication. A third participant (Victor) is
not needed, as Alice assumes Victor's role. We name the photon pairs
2\&3 and 1\&4 \textquotedblleft{}entanglement groups\textquotedblright{}
(EG). Without any loss of generality, we study the case with \emph{two}
particles per entanglement group, while future applications with more
particles per EG are imaginable. Together we will call those two EGs
at Alice and Bob an \textquotedblleft{}EG pair\textquotedblright{}.
Figure \vref{fig:basic-concept-of EMG} shows the EG concept schematically.
The EG concept is the central architectural component of the PSBA
procedure.

Alice acts as the sender, Bob as the receiver. In order to send a
single bit, Alice decides how she will measure her photons 2 and 3:
Either she performs a Bell-state measurement (BSM) and thus she entangles
photons 1 and 4 at Bob (Entanglement Swapping), or she measures her
photons 2 and 3 individually, which will not entangle photons 1 and
4 with each other. However, the BSM could bring the photon pairs 2\&3
and 1\&4 unpredictably into one of four Bell states. Therefore, purposefully
setting the type of the Bell state cannot be used for encoding information.
All we seem to know is that entanglement swapping brings 2\&3 and
1\&4 into exactly the $same$ unknown Bell state \citep{Pan.1998,Jennewein.2001}.
We can not use the type of swapped Bell states for encoding, but we
can use the fact that the photons 1\&4 are \emph{in one of the four
Bell states (it does not matter which one)} for encoding. If we could
detect that the photon pair 1\&4 is probably entangled, this would
be sufficient for encoding. In the meantime, there are several efficient
technical approaches for detecting even more than two of the four
Bell states on an analyzed photon pair \citep{Schmid.2009,Weihs.2001}.
In particular, the fermionic behavior of two photons 1 and 4 in the
Bell state 

\medskip{}

$\Psi_{14}^{-}=\frac{1}{\sqrt{2}}(\Ket{H}\Ket{V}-\Ket{V}\Ket{H})$

\begin{figure}
\includegraphics[width=1\columnwidth]{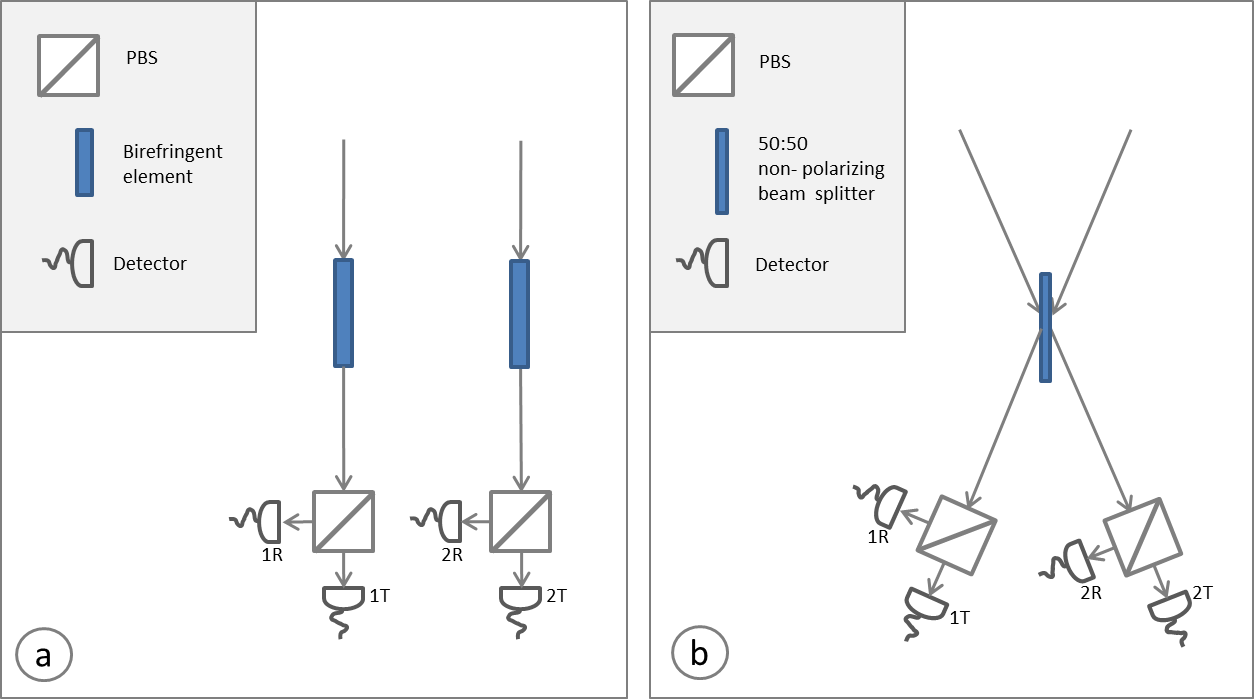}

\caption{a: Statistical distinction of either maximally polarization-entangled
photons or unentangled photons: For $\Psi^{\pm}$ states, the H and
V parts of the wave function will be separated by strongly birefringent
material (as used in \citep{Kwiat.1998}). Therefore the photons will
be detected as time-separated. Photon pairs in $\Phi^{\pm}$ (with
no detection-time delay) should show identical behavior at the beam
splitters, while photon pairs in $\Psi^{\pm}$ should show exactly
the opposite behavior. Uncorrelated photons (time-separated or not)
should show no quantum correlation at the polarizing beam splitters.
b: statistical detection of $\Psi^{-}$ at a 50:50 non-polarizing
beam splitter. Only $\Psi^{-}$ entangled photons will be detected
in different output ports of the non-polarizing beam splitter, while
photons in one of the other three Bell states will end up in the same
output port. For unentangled (distinguishable) photons we probably
will not see the typical Hong-Ou-Mandel-Dip \citep{Hong.1987}. Of
course, for unentangled and \emph{indistinguishable} photons the author
assumes a perfect HOM-Dip: Under perfect HOM conditions not one unentangled
photon pair would be able to let its two photons emerge from different
output ports of the 50:50 non-polarizing beam-splitter, while about
25 percent (swapped $\Psi^{-}$states) of the entangled pairs will
be able to let their two photons emerge from different output ports.
For a general discussion of the possible analysis results see Figure
\vref{fig:Expected-(idealized)-matrix} \label{fig:Statistical-distinction-of}}
\end{figure}

\medskip{}

at a beam splitter as opposed to the bosonic behavior of the other
three Bell states would be of great benefit. Two probably possible
setups for distinguishing a sequence of maximally polarization-entangled
photon pairs from a sequence of unentangled photon pairs are shown
in Figure \vref{fig:Statistical-distinction-of}. One of them (setup
\emph{b}) uses the said detection of the Bell state $\Psi^{-}$(as
a simplified version of setup \emph{b,} shown in Figure \vref{fig:Expected-(idealized)-matrix}).

As described above, with just one photon pair Bob will not be able
to decide with sufficient certainty whether Alice has performed BSM
or SSM. But a statistical evaluation of correlation tests (as described
in Figure \vref{fig:Statistical-distinction-of}) on a larger sequence
of EGs could help Bob out: As described above, the author assumes
that Bob can count coincidences and can construct one of two characteristic
correlation diagrams: A characteristic and significant mixture of
detected correlations (we are not necessarily interested in identifying
particular Bell states) will lead to the certainty Bob needs to determine
whether Alice has performed multiple BSMs or multiple SSMs.

\begin{figure}
\includegraphics[width=1\columnwidth]{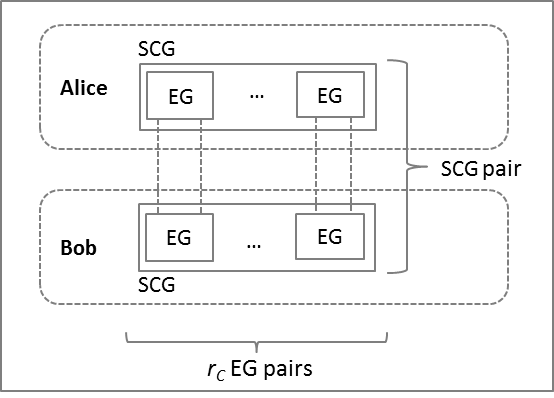}

\caption{SCG concept: One SCG pair contains $r_{C}$ EG pairs in sequence.
To send the value 1b, Alice performs a Bell-state measurement on all
photon pairs (EGs) in her SCG; to send the value 0b, Alice measures
all of her photon pairs individually. \label{fig:SCG-concept:-One}}
\end{figure}

Therefore, in preparation for their PSBA communication, Alice and
Bob share a \textquoteleft{}sufficiently' large number of EG pairs
with each other and pay attention to the same order of the EGs. Alice
and Bob each henceforth have a sequence of EGs (EG1, EG2, ...) at
their disposal. Furthermore, Alice and Bob have to determine a parameter\emph{
r\textsubscript{C }}($r$ for \textquoteleft{}reliability', $C$
for \textquoteleft{}channel') for their PSBA channel by testing: How
many photon pairs (EGs in sequence) have to be entangled by BSM at
Alice before Bob can detect the transmitted value 1b with sufficient
certainty? (Alice and Bob stipulate a concrete probability $p$ for
the term \textquoteleft{}sufficient certainty'.) The value of \emph{r\textsubscript{C }}may
be high as long as it is finite. After determining\emph{ r\textsubscript{C }},
both Alice and Bob partition their EG sequence in EG blocks with \emph{r\textsubscript{C }}EGs
each. We refer to those blocks as SCGs (statistical correction groups).
Figure \vref{fig:SCG-concept:-One} shows the SCG concept schematically. 

\begin{figure}
\includegraphics[width=1\columnwidth]{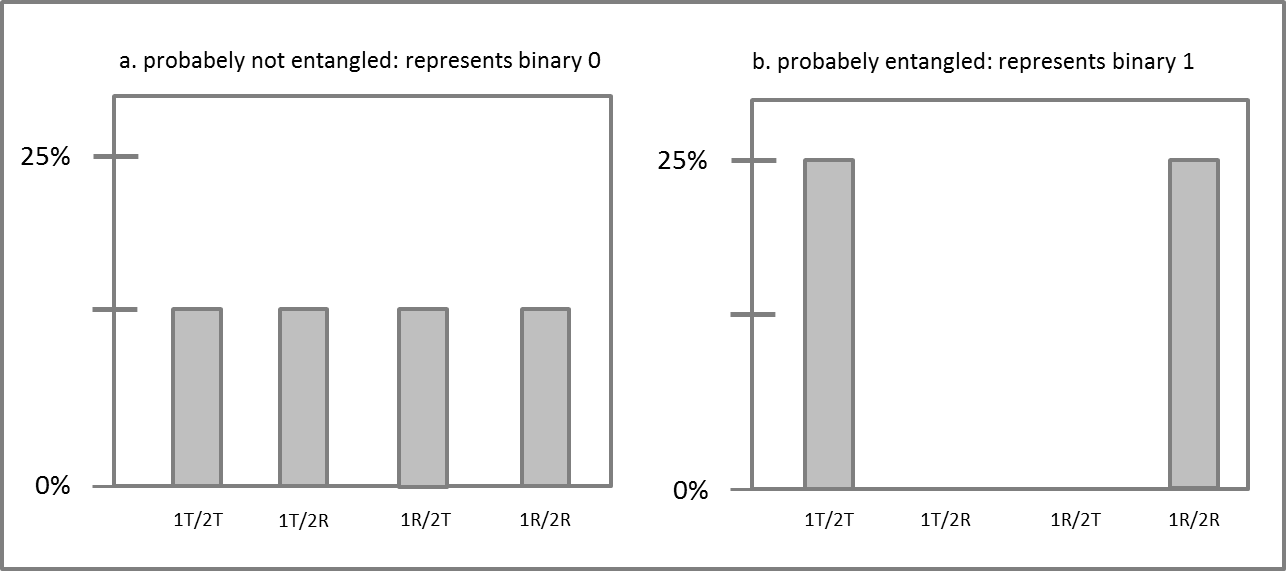}

\caption{Correlation diagrams (theoretical/ideal results of the gedankenexperiment
for the analyzer-setup shown in part (a) of Figure \vref{fig:Statistical-distinction-of}):
The diagrams here show the correlations for $\Phi^{\pm}$ only, but
- as described above - Bob can distinguish $\Phi^{\pm}$and $\Psi^{\pm}$
by time-delayed detections. So Bob can generate these diagrams without
additional information from Alice. If all photon pairs in Bob's SCG
are unentangled (and randomly polarized), Bob will find a correlation
diagram similar to \emph{a} (left). If all photon pairs are entangled
in one of the four Bell states, Bob will find (for $\Phi^{\pm}$)
a correlation diagram similar to \emph{b} (right). For the analyzer-setup
shown in part (b) of Figure \vref{fig:Statistical-distinction-of}:
Finally, the author assumes a single non-polarizing 50:50 beam splitter
as a sufficient setup for detecting a stream of (by entanglement swapping)
entangled photon pairs (see Figure \vref{fig:Expected-(idealized)-matrix}
for the discussion of this setup)\label{fig:ideal-results}}
\end{figure}

\begin{figure}
\includegraphics[width=1\columnwidth]{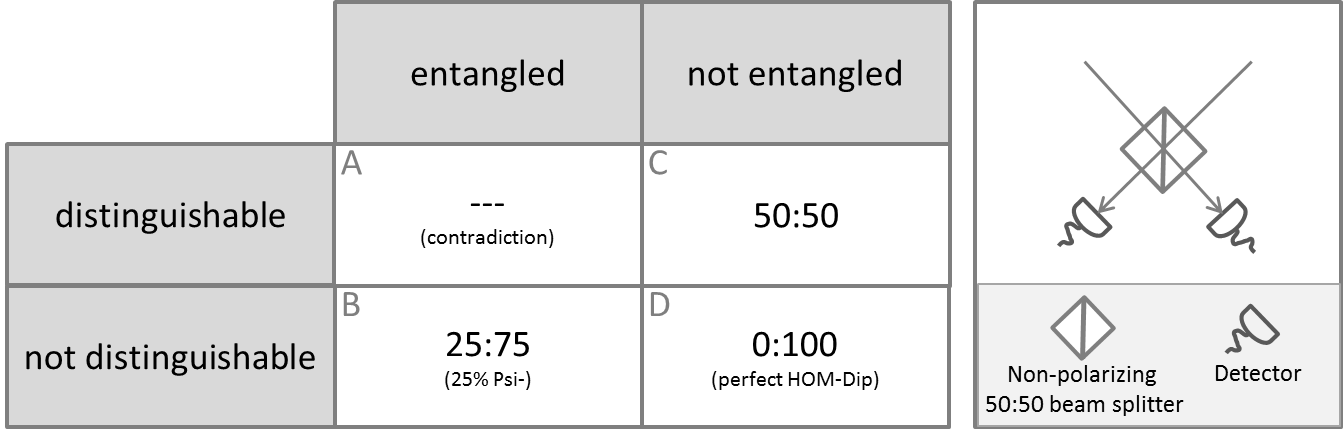}

\caption{The matrix of possible correlation proportions (left) for a simplified
Bell-state analyzer setup (right): In order to allow indistinguishability,
for the possible results (shown in the matrix) an identical time of
arrival for the two photons at the beam splitter is assumed. The left
part of each of the four given correlations proportions represents
the statistically expected percentage of photon pairs with its photons
detected at different exits of the beam splitter, the right part of
each proportion represents the expected percentage of photon pairs
with both photons leaving the beam splitter through the same output
port. As described before, the swapped Bell-states (Alice sends 1b)
will produce 25 percent of ``two-exits-pairs'' (by the $\Psi^{-}$
states) and 75 percent of ``one-exit-pairs'' (case B) in the analysis.
Case A should be a contradiction under the assumption of the same
time of arrival at the beam splitter. When Alice sends 0b (and performs
no entanglements swappings) the photons in the photon pairs in Bob's
SCG are expected to be distinguishable (by polarization and/or frequency,
case C). Therefore, Bob's analysis would generate a 50:50 proportion.
The other (theoretically possible) case (D): All photon pairs in Bob's
SCG are not entangled but indistinguishable (How this could be realized
technically by Bob in a PSBA scenario seems to be questionable). In
this case Bob could probably see a (theoretically perfect) HOM dip.
However, in both 'unentangled cases' (C and D) the proportion would
be significat different from the proportion 25:75 (produced by swapped
Bell-states). Therefore, Bob can distinguish a SCG with photon pairs
in swapped Bell-states from a SCG with unentangled photon pairs. \label{fig:Expected-(idealized)-matrix}}
\end{figure}

Consequently, Alice and Bob now each have a sequence of SCGs at their
disposal. In order to transmit a single bit (0b or 1b), Alice and
Bob use one SCG pair per bit. For the next bit they use the next unused
SCG pair in the SCG sequence. Each SCG pair can be used only once
for a bit transmission, since after entanglement swapping (or SSM)
the entanglements between Alice and Bob (as shown in Figure \vref{fig:SCG-concept:-One})
will be dissolved. In order to send the binary value 1b, Alice executes
a Bell-state measurement on every single photon pair (EG) of her current
SCG; to send the value 0b, Alice performs separable-state measurements
analogically. As a consequence of Alice's Bell-state measurements,
every single EG (in the ideal case) in Bob's current SCG will be in
a Bell state. Bob executes a correlation analysis (as described above)
on every photon pair of his SCG. After doing so, Bob can determine
with a sufficient probability $p<1$ whether his photon pairs (1\&4)
in this SCG were more likely to be unentangled (Bob reads 0b) or entangled
with each other (Bob reads 1b). For the setup shown in part (a) of
Figure \vref{fig:Statistical-distinction-of} the author expects (under
theoretical/ideal conditions) two correlation diagrams similar to
those in Figure \vref{fig:ideal-results}.

Hence Alice can transmit binary-encoded messages to Bob with sufficient
certainty by using PSBA. For example, if Alice wants to send the message
\textquotedblleft{}FASTER\textquotedblright{} (in 8-bit ASCII Alice
needs 6 bytes or 48 bits) to Bob, she uses 48 SCGs to encode this
message. Bob should know the length of the message beforehand (i.e.
he should know how many SCGs he has to analyze). Therefore, e.g. one
byte (and therefore 8 SCGs) in front of the message data could be
used as such a length field. In addition, before sending (encrypted)
messages photons could be used for security by Quantum Key Distribution
\citep{Bennett.1984} (see discussion). As described above, for this
transmission neither classical ways of communication were used in
addition to quantum mechanisms, nor were (entangled) particles (photons)
sent from Alice to Bob (or vice-versa) after information encoding.
As the central consequence, the \textquoteleft{}instantaneous' effect
of entanglement swapping in combination with statistical Bell-state
detection seems to allow communication faster than light.

One important question still has to be discussed in order to complete
the PSBA procedure: How does Bob know that Alice has sent him a message?
In other words: When does Bob know that he can analyze the photon
pairs of the next SCG for reading the next bit? A first, surely practicable
solution is a fixed time interval Alice and Bob stipulated for synchronization
purposes. After each time interval Bob reads one SCG (the term \textquotedblleft{}polling\textquotedblright{}
could be derived from computer science here) regardless of whether
Alice has sent anything. If this SCG represents a set bit (1b), this
could be interpreted as a previously agreed indicator for more transmitted
data (following in the next SCGs). By putting more than two photons
in an EG (and swapping the entanglement to different combinations
of them), even the synchronization of this time interval could be
realized. This approach probably works; however, it is not especially
elegant, as potentially unnecessarily SCGs have to be consumed. For
growing classical transmission distances (light weeks, light months
or more) such a polling frequency can be slowed down proportionally
while preserving the benefit of a faster transmission without heavily
loading Alice\textquoteright{}s and Bob's SCG pools.

Another, second variant would be more efficient (as no SCGs were consumed
quasi-uselessly), but - if technically correct and realizable - this
approach would be revolutionary. If we trust the delayed-choice experiments
of \citep{XiaosongMa.2012} and \citep{Megidish.2013}, and interpret
them correctly, Bob would not have to know when Alice sends data,
but could analyze the photon pairs of one of his SCGs correctly at
\emph{any} point of time (as described above without additional (measurement)
information from Alice), since the said experiments seem to show that
he can analyze his photon pairs (1\&4) correctly even \emph{before}
Alice executes BSM (or separable-state measurements) on her SCG's
photon pairs. Hence, timing between Alice and Bob would no longer
be necessary and Bob could read a new message starting at the next
SCG whenever he tries, as long as Alice uses this SCG for transmitting
a message at any point in the future. The author has doubts about
whether this would be possible, but if it were, this would lead to
the usual discussions about temporal paradoxes, since Bob could then
read a message before Alice sends it.

\section{Further Discussion And Conclusion }

\begin{figure}
\includegraphics[width=1\columnwidth]{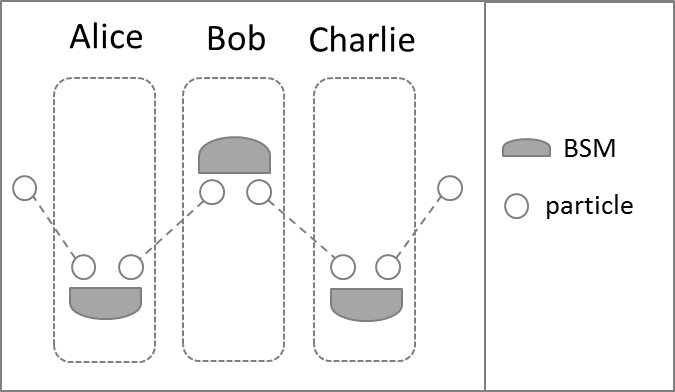}\caption{Concept of concatenated entanglement swapping protocol as described
in \citep{Khalique.2013}. Two photons of two entangled photon pairs
are, in contrast to PSBA, spatially separated in an initial stage.
With this approach a particle's state could be teleported over multiple
hops and long distances. As for quantum teleportation, classical communication
will be required to reconstruct the correct transferred states. \label{fig:concated-ESC}}
\end{figure}

\begin{figure}
\includegraphics[width=1\columnwidth]{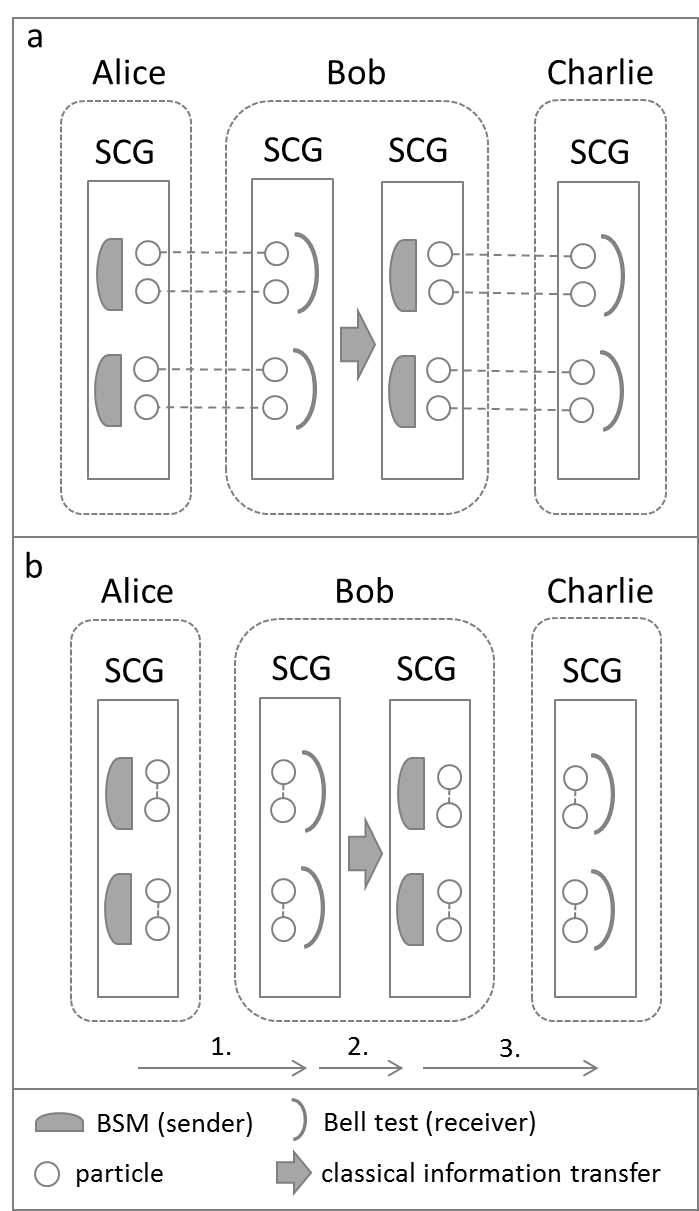}

\caption{Multi-hop PSBA communication example. Without any loss of generality
SCGs with \emph{two} EGs were shown here as an example. Bob acts as
a repeater in this scenario. The direction of transmission is from
the left to the right (a): The initial state is shown: No entanglement
swappings have been performed so far. (b): Transmission of the binary
value 1b. Alice performs Bell-state measurements on all photon pairs
in her current SCG (1.), which leads to entangled photons (2\&3) at
Alice and also at Bob\textquoteright{}s photon pair (1\&4). Bob detects
entanglements in his SCG and interprets this as a binary 1. Bob forwards
this information by executing Bell-state measurements on an SCG which
was entangled with another SCG at Charlie (3.). The communication
using PSBA is limited to one transport \textquoteleft{}hop\textquoteright{}.
Therefore after each hop the transmitted information has to be decoded
from one SCG, transferred on classical ways to another SCG, and be
re-encoded into the sending SCG of the next PSBA transport segment
(Bob \textendash{} Charlie). This is a local operation and as such
it would take milliseconds on each repeater node in a multi-hop PSBA
communication. For larger spatial transmission distances these milliseconds
will have hardly any influence on the overall \textquoteleft{}instantaneous\textquoteright{}
transmission. \label{fig:executed-ESC-multi-hop-communication}}
\end{figure}

The concept of PSBA can be discussed in relation to several other
technical approaches. The goal of quantum teleportation, for example,
is to transfer one or more states from one particle to another particle.
Several experiments like \citep{Ma.2012} have shown quantum teleportation
over distances of more than 140 kilometers so far. Quantum teleportation
needs additional information (via classical communication channels)
to reconstruct the correct teleported states. Quantum teleportation
represents information by a particle\textquoteright{}s states and
uses entanglement for transportation. PSBA is different, because PSBA
represents information by an entanglement and uses entanglement swapping
for transportation. PSBA is also different, because PSBA does not
need any additional classical information transport. As PSBA is a
statistical approach (i.e. it uses a large number of particles), it
has a lower efficiency regarding the density of encoded information
(bits per particle) than quantum teleportation. Other approaches \citep{Khalique.2013}
use (concatenated) entanglement swapping for communication, but (similar
to quantum teleportation) these approaches transport a particle's
state over one (or more) entanglement swapping \textquoteleft{}hops',
while a PSBA communication path is built to be just one \textquoteleft{}hop'
long, since photons 1 and 4 are at the \emph{same} communication participant.
As other quantum teleportation approaches the approach proposed in
\citep{Khalique.2013} will also need additional classical communication
to reconstruct the correct transferred state. Nevertheless, multi-hop
PSBA communication is possible: Between two hops (at a \textquoteleft{}repeater')
the transmitted data has to be decoded and \textquoteleft{}classically'
transferred into another SCG of the next segment of the transmission
route. Figure \ref{fig:concated-ESC} shows the concept of concatenated
entanglement swapping as described in \citep{Khalique.2013}. Figure
\vref{fig:executed-ESC-multi-hop-communication} shows a multi-hop
PSBA transmission of a binary 1 from Alice to Charlie.

PSBA also differs from approaches like Dense Coding \citep{Bennett.1992,Mattle.1996}),
since for Dense Coding either a particle as one of two necessary carriers
of \textquoteleft{}parts of information' has to be sent to the receiver
via classical ways, or measurement results have to be exchanged on
classical ways. When using PSBA, entangled photons are distributed
on classical ways as well, but no particles or measurement results
are transmitted spatially \emph{either during or after} encoding any
information. For the same reason, PSBA differs from quantum secure
direct communication protocols (QSDC) like \citep{Bostrom.2002,HyungJin.2010}.
The focus of PSBA is the speed of transmission, so it would not be
helpful to exchange measurement results on classical ways while communicating
just for the purpose of security. As one can easily see, PSBA is compatible
with Quantum Key Distribution (QKD) \citep{Bennett.1984}, as this
encryption key will be generated (by using classical channels) before
sending (encrypted) messages using PSBA. To realize QKD, Alice and
Bob can take photon pairs from their EG pairs. The QSCD protocol presented
in \citep{HyungJin.2010} is thus similar to PSBA, since Alice and
Bob have shared two entangled particle pairs as well, but \emph{both}
Alice and Bob perform Bell-state measurements on their photons pairs
(2\&3), (1\&4). In clear contrast to PSBA, in \citep{HyungJin.2010}
the measurement results have to be announced on a classical way in
order to enable Alice and Bob to decode the received messages. 

\medskip{}

Within this article a procedure called \textquotedblleft{}Probabilistic
Swapped-Bell-States Analysis\textquotedblright{} is proposed. PSBA
seems to allow the transmission of information at a speed faster than
light. To realize this, entanglement swapping is used as a transport
channel where statistical detection of (swapped) entanglements is
used for transmitting binary data. The PSBA approach has been discussed
in relation to several existing approaches, such as quantum teleportation,
\textquoteleft{}classical' (concatenated) entanglement swapping protocols,
dense coding approaches as well as QKD and QSCD. An approach for multi-hop
PSBA communication has been described. If PSBA mechanism should work,
this would contradict essential parts of Albert Einstein's theory
of relativity. In that case, as the author is not a quantum expert,
these specialists will derive the consequences and show how PSBA can
be implemented in the most efficient manner. The author has profound
doubts about the correctness of the PSBA mechanism, since everything
needed (mechanisms for detecting Bell states as well as the mechanism
of entanglement swapping) has been known for more than 15 years; the
author considers it improbable that not one quantum expert could have
seen this possibility in this time. Hence the PSBA mechanism should
be either incorrect or at least impracticable.

\medskip{}

The author must thank several professors and PhD graduates of theoretical
physics and optics who have invested time in answering the author's
questions. The individual acknowledgements would follow in an update
of this publication in case of confirmed relevance of this paper.

\renewcommand{\labelenumi}{(\roman{enumi})}

\bibliographystyle{elsarticle-num}
\bibliography{Nachrichtenuebertragung}

\end{document}